\documentclass[twocolumn]{aastex61}
\pdfoutput=1 
\usepackage{amsmath,amstext}
\usepackage[T1]{fontenc}
\usepackage{apjfonts} 
\usepackage{hyperref}
\usepackage[figure,figure*]{hypcap}
\usepackage{color}

\usepackage{soul}

\shorttitle{Synthetic Radio Imaging}
\shortauthors{S.P. Moschou et al.}

\begin{document}

\title{Synthetic Radio Imaging for Quiescent and CME-flare Scenarios}

\author{Sofia-Paraskevi Moschou}
\affiliation{Harvard-Smithsonian Center for Astrophysics, 60 Garden Street, Cambridge MA 02138, USA}

\author{Igor Sokolov}
\affiliation{Center for Space Environment Modeling, University of Michigan, Ann Arbor, MI 48109-2143, USA}

\author{Ofer Cohen}
\affiliation{Lowell Center for Space Science and Technology, University of Massachusetts, 600 Suffolk St., Lowell, MA 01854, USA}

\author{Jeremy J. Drake}
\affiliation{Harvard-Smithsonian Center for Astrophysics, 60 Garden Street, Cambridge MA 02138, USA}

\author{Dmitry Borovikov}
\affiliation{University of New Hampshire Main Campus, Durham, NH 03824, USA}

\author{Justin C. Kasper}
\affiliation{Center for Space Environment Modeling, University of Michigan, Ann Arbor, MI 48109-2143, USA}

\author{Julian D. Alvarado-Gomez}  
\affiliation{Harvard-Smithsonian Center for Astrophysics, 60 Garden Street, Cambridge MA 02138, USA}

\author{Cecilia Garraffo}
\affiliation{Harvard-Smithsonian Center for Astrophysics, 60 Garden Street, Cambridge MA 02138, USA}
\affiliation{IACS, Harvard University, 33 Oxford Street Cambridge, MA 02138, USA}

\begin{abstract}
Radio observations grant access to a wide range of physical processes through different emission mechanisms. These processes range from thermal and quiescent to eruptive phenomena, such as shock waves and particle beams. We present a new synthetic radio imaging tool that calculates and visualizes the Bremsstrahlung radio emission. This tool works concurrently with state-of-the-art Magnetohydrodynamic (MHD) simulations of the solar corona using the code BATS-R-US. Our model produces results that are in good agreement with both high and low frequency observations of the solar disk. In this study, a ray-tracing algorithm is used and the radio intensity is computed along the actual curved ray trajectories. We illustrate the importance of refraction in locating the radio emitting source by comparison of the radio imaging illustrations when the line-of-sight instead of the refracted paths are considered. We are planning to incorporate non-thermal radio emission mechanisms in a future version of the radio imaging tool.

\end{abstract}

\keywords{Sun: sunspots, coronal mass ejections (CMEs), flares, activity, chromosphere, radio radiation}

\section{Introduction}
\label{S:1}

Radio electromagnetic wavelengths have a favorable atmospheric window for ground observations. A number of very diverse radio emission mechanisms that are generated by phenomena that range from quiescent and thermal processes to eruptive events, such as shocks and particle acceleration due to reconnection, can be observed from the Earth {\citep[e.g.][]{Dulk:85}}. These characteristics make the incorporation of synthetic radio tools into solar and stellar coronal simulations of great value for understanding and disentangling the complicated underlying emitting physical processes. Such tools can help to provide the observational community with the ability to interpret observations in terms of detailed parameterized physical models, with detection limits for each underlying radio-generating physical process, and can help in designing future Earth- and space-based facilities. 

The Sun is of special importance, not only because it is a great laboratory for plasma physical mechanisms, but also due to its proximity, which makes it possible for radio observers to spatially resolve the structures above the photosphere \cite[e.g.][]{Oberoi.etal:11,Mohan.Oberoi:17}.
{The morphology of the radio emissions from the quiet solar atmosphere depends on the magnetohydrodynamic (MHD) plasma quantities (density, temperature and magnetic field) as well as the radio emission mechanisms. The quiescent solar disk appears to have decreasing size and better resolved fine-scale structures with increasing observational frequency \citep[e.g.][]{Mercier.Chambe:09,Mercier.Chambe:15}. As explained by \cite{Lee.etal:99}, higher frequencies are able to penetrate deeper in the solar atmosphere and have access to stronger magnetic fields and thus smaller scale structures. Radio observations have been used extensively for the study of quiet Sun features such as prominences and cavities \citep{Dulk.Sheridan:74,Shevgaonkar.etal:88,Marque:04}.}
Gyroresonance and Bremsstrahlung observations have been used for quantifications of the global coronal magnetic field in the Sun~\citep{Casini.etal:17}. \cite{Casini.etal:17} discuss ways that observations can constrain and confirm computational models of global stellar magnetism.

{In addition to quiescent conditions, solar radio observations can provide essential information about events giving rise to flares, coronal mass ejections (CMEs) and solar energetic particles (SEPs) \citep[e.g.][]{Gopalswamy.Kundu:92,Reiner.Kaiser:99,Kouloumvakos.etal:14,Chen.etal:15,Winter.Ledbetter:15}. The multi-mechanism radio emission in the solar atmosphere calls for the development of sophisticated MHD models that capture the complicated nature of the different intertwined radio emissions. In their paper, \cite{Chen.etal:15} have, for the first time, directly observed the termination shock below a reconnection site, which was then simulated using MHD models. A fully analytic formalism accounting for electron beams, Langmuir waves, and radio emission was developed in 
\cite{Schmidt.Cairns:12} and was then used in combination with MHD models to capture Type II radio bursts in the solar corona \citep{Schmidt.Cairns:13,Schmidt.Cairns:16}.
}

{Even though the quiescent solar atmosphere has been imaged in a range of wavelengths \citep[e.g.][]{Mercier.Chambe:09,Oberoi.etal:11,Mercier.Chambe:15,Mohan.Oberoi:17} and radio bursts have been studied extensively for several decades already \citep[e.g.][]{Gopalswamy.Kundu:92,Reiner.Kaiser:99,Dougherty.etal:02,Kouloumvakos.etal:14,Winter.Ledbetter:15}, CMEs are difficult to image in radio frequencies as they become diffuse soon after the eruption. For that reason, only a few studies were able to image radio CMEs and in particular their eruptive phase and early development. The initiation and evolution of CMEs has been observed in radio frequencies of the order of a few hundreds of MHz \citep{Bastian.etal:01,Maia.etal:07,Demoulin.etal:2012}. The emission was attributed to incoherent synchrotron emission of electrons with energies in the MeV range as they interact with the local magnetic field. More recently, \cite{Zucca.etal:18} were able to produce a three-dimensional reconstruction of a CME using Low-Frequency Array (\textit{LOFAR}) radio observations.}

We have developed and incorporated into our numerical simulations of coronae and winds a synthetic radio imaging tool that captures Bremsstrahlung emission along the actual curved paths of each Radio-ray. {Among the abovementioned diverse radio emission mechanisms, we focus on the Bremsstrahlung emission, due to the relative simplicity of its implementation in global MHD models. Although more complex emission mechanisms are not yet addressed by this study, the present work can guide future algorithm upgrades.} The main focus of this paper is to demonstrate the new capabilities of the radio synthetic imaging algorithm and present the first qualitative comparisons to observations. In Section \ref{sec:method}, we elaborate on the schemes used, while in Section \ref{sec:results} we present our results and compare with observations. Finally, we finish with a short discussion and our conclusions in Sections \ref{sec:discussion} and \ref{sec:conclusions}. 

\section{Computational Method} \label{sec:method}

The basis of our synthetic radio emission and refraction tool is a numerical model describing the plasma and magnetic field conditions of the solar corona, wind and coronal mass ejections within which the radio signal originates. 
Numerical simulations of the solar corona and wind are performed using the state-of-the-art Block-Adaptive Tree Solarwind Roe Upwind Scheme (BATS-R-US) code \citep[see e.g.][]{Powell.etal:99,Toth.etal:12,vanderHolst.etal.14}.
We run data-driven single-fluid MHD simulations using synoptic (time-averaged) magnetograms over the period of a full Carrington Rotation (CR) for the photospheric magnetic field. For the purposes of this study we are using \textit{GONG}\footnote{\textit{GONG} magnetograms are obtained by a network of ground based telescopes and are available  at \url{https://gong.nso.edu/data/magmap/}.} magnetograms as input for our simulations. To accelerate and heat the solar wind we are using a scheme based on Alfv{\'e}n wave turbulence \citep[AWSoM-R - see][]{Sokolov.etal.13,Oran.etal.13,vanderHolst.etal.14}, while for the CME eruption and propagation we make use of the Eruptive Event Generator based on the eruptive flux-rope \cite{Gibson.Low:98} (GL98) model \citep[EEGGL - see][]{Jin.etal:17a,Jin.etal:17b,Borovikov.etal.17}. These models are parts of the Space Weather Modeling Framework \citep[SWMF - see][]{Toth.etal:12}. 

Here, we simulate the CME event of 7 March 2011 at 19:40 that occurred during Carrington Rotation (CR) 2107. We chose this event because it was analyzed in earlier computational studies that used BATS-R-US \citep[e.g.][]{vanderHolst.etal.14,Jin.etal:17b}. The GL98 eruptive flux-rope parameters can be found in Table \ref{tab1}. 

\begin{deluxetable}{lcc}[h!]
\tablecaption{Flux rope parameters of the GL98 CME model used for CR 2107.\label{tab1}}
\tablecolumns{3}
\tablenum{1}
\tablewidth{2pt}
\tablehead{\colhead{Parameter} & \colhead{Unit} & \colhead{Value}}
\startdata
Latitude & deg & 27.5 \\
Longitude & deg & 157.5 \\
Orientation\tablenotemark{i} & deg & 128.45 \\
Stretch ($a$) & \nodata & 0.6 \\
Pre-stretch distance ($r_1$) & $R_{*}$ & 1.8 \\
Size ($r_0$) &  $R_{*}$ & 0.55 \\
Magnetic strength ($a_1$) & G $R_*^{-2}$ & 8.6\\
Flux rope helicity & \nodata & Sinistral (-) \\
\enddata
\tablenotetext{i}{Measured with respect to the stellar equator in the clock-wise direction.}
\end{deluxetable}

Our radio synthetic imaging tool is based on {the ray-tracing algorithm presented in \cite{Benkevitch:10} and }\cite{Benkevitch.etal:12}, which has been updated for the purposes of this study. Major improvements include: a) the generalization of the way the radio intensity is calculated for different emission mechanisms on Adaptive Mesh Refinement (AMR) grids in arbitrary coordinates (spherical, Cartesian etc), b) the implementation of free-free Bremsstrahlung radiation in addition to the originally implemented (and from now on referred to as) "simplistic" emission mechanism (see below) and c) improving the interpolation scheme for AMR grids \citep{Borovikov.etal.2015}. 

{For the construction of the radio synthetic images in a wide range of frequencies we are using the ray tracing algorithm described in \cite{Benkevitch:10}. A typical radio image of the Sun consists of on the order of thousands to millions of pixels. The intensity of each individual pixel is calculated as an integral along a single ray path. Heliospheric plasmas are highly non-uniform and thus, contrary to the rest of the spectrum, rays with frequencies of the order of a few GHz and lower cannot be considered as straight lines. Efficient ray tracing algorithms need to be used to obtain the individual ray trajectories. The ray tracing algorithm implemented in BATS-R-US calculates radio ray trajectories in space plasmas with known MHD plasma characteristics.}

\subsection{{Refraction Implementation}} \label{subsec:refimpl}

Refraction takes place throughout the entire electromagnetic spectrum.
Radio waves experience the effects of refraction to a greater extent than the rest of the electromagnetic spectrum due to their strongly varying refractive index between media of different densities at typical densities encountered in the outer solar atmosphere. The role of refraction is crucial for accurately locating the source of the radio signal and can lead to great spatial errors if not accounted for. The effects of refraction cannot be ignored for radio waves, since locating the source of each radio producing mechanism is key in advancing our physical understanding of the complex processes involved, such as particle acceleration.

Refraction via calculation of the actual curved path of the radio rays through the computational domain is performed concurrently with our simulation, using the ray-tracing algorithm developed by and described in greater detail in \citet{Benkevitch:10} \citep[see also][]{Benkevitch.etal:12}. A path integration is performed over the ray trajectories to calculate the radio intensity in media with known density distribution. The method uses and quantifies the gradient of the dielectric permittivity, which is a function of the density and the density gradient, both of which are known quantities in our realistic numerical coronal simulations. Each ray trajectory is treated as a curve with position vector $\mathbf{r}$ in the three-dimensional space. Once the index of refraction $n(\mathbf{r},\omega)$ is known in our domain of interest, for a specific frequency $\omega$, the full trajectory can be uniquely determined for a starting point $\mathbf{r_0}$ and a given initial direction vector $\mathbf{v_0}$. By defining the direction vector of the ray trajectory as the first derivative of the positional vector $\mathbf{v}(s)=\mathbf{r'}(s)$, where $s$ the curve arch length, it was shown by \citet{Benkevitch:10} that the density distribution is sufficient to calculate the refraction index for any wave with a given frequency $\omega$.

In Gaussian (CGS) units, the refractive index $n$ is a function $n=\sqrt[]{\epsilon}$ of the dielectric permittivity $\epsilon$, and for isotropic media can be written as 
\begin{equation}
n^2=1-\frac{\omega_p^2}{\omega^2} {.}
\label{e:ref_index}
\end{equation}
Then, assuming quasi-neutrality the hydrogen plasma density is $\rho=m_p n_e$, where $m_p$ is the proton mass. Finally, the dielectric permittivity becomes 
\begin{equation}
\epsilon=1-\frac{\rho}{\rho_{cr}} \,
\end{equation}
with $\rho_{cr}$ being the critical plasma density
\begin{equation}
\rho_{cr}=\frac{m_p m_e \omega^2}{4\pi e^2} \,
\end{equation}
where both the permitivity and the refractive index become zero for a specified frequency $\omega$. Rays with frequency $\omega$ cannot penetrate deeper into the solar atmosphere than the critical surface where the density is larger than the critical density $\rho>\rho_{cr}$ and thus plasma frequency is larger than the ray frequency $\omega_p >\omega$, as for those regions the refractive index becomes imaginary.  For $\omega>\omega_p$, it can be seen from Equation~\ref{e:ref_index} that the refractive index increases with decreasing frequency, such that refraction is more important for low radio frequencies than high frequencies. {Electromagnetic rays that approach the critical surfaces undergo strong refraction and bend away from the surface. For the limit cases this effect resembles reflection.}


\subsection{{Ray Tracing Algorithm}}\label{subsec:raytracing}

{Fermat's principle is used for the calculation of the ray trajectories. The principle states that an electromagnetic ray connecting two points in space will follow a trajectory that minimizes the travel time between starting point A and final point B. The travel time can be written as:
\begin{equation}
T_{A\rightarrow B}=\frac{1}{c}\int^B_A n(\mathbf{r}) \mathrm{d}s .
\end{equation}
It can then be shown that the ray will travel over a path that minimizes the integral of the refractive index, $n$. This means that the variation of this integral over the chosen path is zero:
\begin{equation}
\delta \int^B_A n(\mathbf{r}) \mathrm{d}s = 0 .
\end{equation}
In order to find the vector differential equation of the trajectory defined by Fermat's principle, Euler's equation is used which writes:
\begin{equation}
\mathrm{\frac{d}{d\tau} \left( \frac{d\mathcal{L}}{d\mathbf{\dot{r}}}\right) - \frac{d\mathcal{L}}{d\mathbf{r}}}= 0 ,
\end{equation}
with $\mathcal{L}(\mathbf{r,\dot{r}},\tau)=n(\mathbf{r})\ \sqrt[]{\mathrm{\dot{r}^2}}$ being the Lagrangian, where $\tau$ is an independent time variable. After some mathematical manipulations and by introducing the direction vector $\mathrm{\mathbf{v}}=\mathrm{d\mathbf{r}/ds}$ we get a set of six first-order equations
\begin{equation}
\label{eq:diff6}
\begin{cases}
\frac{\mathrm{d\mathbf{r}}}{\mathrm{d}s}=\mathrm{\mathbf{v}} \\
\frac{\mathrm{d\mathbf{v}}}{\mathrm{d}s}=\mathrm{\mathbf{v}}\times \left(\frac{\nabla n}{n}\times \mathrm{\mathbf{v}} \right) .
\end{cases}
\end{equation}
This set of differential equations is solved by the ray tracing algorithm. The solution of the equation system~\ref{eq:diff6} is a naturally parametrized curve $\mathbf{r}=\mathbf{r}(s)$, with $s$ being the ray arc length. The dielectric permittivity can be written as $\nabla \epsilon=-\nabla\rho/ \rho_{cr}$. Then the relative gradient of the refractive index $\frac{\nabla n}{n}$ becomes
\begin{equation}
\frac{\nabla n}{n}=\frac{\nabla \epsilon}{2\epsilon}=-\frac{\nabla
\rho}{2(\rho_{cr}-\rho)}.
\end{equation} 
Thus the gradient of the refractive index can be unambiguously determined if we know the plasma density, its gradient and the critical density.}

{A mathematical analogy between the ray equations~\ref{eq:diff6} and the equations for the motion of a particle in a magnetic field is used in our algorithm. For the solution of equation system~\ref{eq:diff6} a second-order difference scheme \citep{Crank.Nicolson:47} is derived using a modified version of the \cite{Boris:70} approach. While the \cite{Boris:70} algorithm ensures particle energy conservation, the ray trajectory computation uses a length conservation law. More specifically, in the original scheme the squared  particle velocity $\mathbf{v}^2$ (particle energy) is being conserved, where $\mathbf{v}(t)=\mathbf{r^\prime}(t)$ . Instead, our scheme uses the natural parametrization of the ray curve to derive and conserve the squared direction vector to unity $\mathbf{v}^2=1$, where $\mathbf{v}(s)=\mathbf{r^\prime}(s)$. For that a vector quantity $\mathbf{\Omega}$ is introduced as 
\begin{equation}
\mathbf{\Omega}=\frac{\nabla n}{n}\times \mathrm{\mathbf{v}}
\end{equation}
and the differential equation~\ref{eq:diff6} is then written as
\begin{equation}
\label{eq:diffnew}
\begin{cases}
\frac{\mathrm{d\mathbf{r}}}{\mathrm{d}s}=\mathrm{\mathbf{v}} \\
\frac{\mathrm{d\mathbf{v}}}{\mathrm{d}s}=\mathrm{\mathbf{v}}\times \mathbf{\Omega} .
\end{cases}
\end{equation}
The curvature of the ray trajectory is $\kappa= |\mathbf{r^{\prime\prime}}(s)|$ and the second equation in \ref{eq:diffnew} writes $\mathbf{r^{\prime\prime}}=\mathbf{r^{\prime}}\times \mathbf{\Omega}$. It then can be shown that the length of the vector $\mathbf{\Omega}$ is equal to the ray trajectory curvature $|\mathbf{\Omega}|=|\mathbf{r^{\prime\prime}}|=\kappa$.}

{A leapfrog difference scheme is used for finding the solution on the half points of the ray trajectory. The second differential equation of the system~\ref{eq:diffnew} can be approximated by
\begin{equation}
\frac{\mathbf{v_1}-\mathbf{v_0}}{\mathrm{d}s}=\mathbf{v}_{1/2}\times \mathbf{\Omega}_{1/2}
\end{equation}
and by using the approximation $\mathbf{v}_{1/2}=(\mathbf{v_1}+\mathbf{v_0})/2$ it leads to the implicit \cite{Crank.Nicolson:47} scheme
\begin{equation}
\mathbf{v_1}-\mathbf{v_0}=(\mathbf{v_1}+\mathbf{v_0})\times \mathbf{\Omega}_{1/2}\frac{\mathrm{d}s}{2} ,
\end{equation}
which ensures unconditional stability. After some mathematical manipulations we arrive at the explicit Boris algorithm known as \textit{CYLRAD} \citep{Boris:70}. So the ray tracing process can be summarized by the following five equations
\begin{align}\label{eq:boris}
& \mathbf{r}_{1/2}=\mathbf{r}_{0}+\mathbf{v}_{0}\frac{\mathrm{d}s}{2} \nonumber \\
& \mathbf{\Omega}_0=\left(\frac{\nabla n}{n} \right)_{1/2}\times \mathbf{v}_{0} \frac{\mathrm{d}s}{2} \nonumber \\
& \mathbf{\Omega}_{1/2}=\left(\frac{\nabla n}{n} \right)_{1/2}\times (\mathbf{v}_{0}+\mathbf{v}_{0}\times \mathbf{\Omega}_0) \frac{\mathrm{d}s}{2} \\
& \mathbf{v_1}=\mathbf{v_0}+\frac{2}{1+\left(\mathbf{\Omega}_{1/2}\frac{\mathrm{d}s}{2}\right)^2}(\mathbf{v}_{0}+\mathbf{v}_{0}\times \mathbf{\Omega}_{1/2}) \times \mathbf{\Omega}_{1/2} \nonumber \\
& \mathbf{r}_{1}=\mathbf{r}_{1/2}+\mathbf{v}_{1}\frac{\mathrm{d}s}{2} \nonumber
\end{align}
which describe the iterative process starting from the point $(\mathbf{r}_{0},\mathbf{v}_{0})$ by which we find the new one $(\mathbf{r}_{1},\mathbf{v}_{1})$.}

{An adaptive step $\mathrm{d}s$ procedure is in place for prediction control and to ensure that rays do not penetrate the region below each respective critical surface ($\rho>\rho_{cr}$), as rays with different frequencies will have critical surfaces with different morphologies. Finally, there is a correctness control subroutine in place that uses a linear scheme to predict the sign of the dielectric permittivity at each next point ($(\mathbf{r}_{1},\mathbf{v}_{1})$) and ensures that no ray crosses the critical surface. For a point beyond the critical surface the dielectric permittivity would be negative. If at the next point the dielectric permittivity is positive the Boris scheme (\ref{eq:boris}) is used and the ray tracing continues normally. However, if the dielectric permittivity is negative at the next point, a) the code goes back and calculates the distance to the critical surface, b) the ray is now approximated by a parabola, and c) the point is then switched to the symmetric point of the parabola. This process is essentially a reflection of the ray at the critical surface, which then travels away from the surface. The linear prediction scheme though could fail due to highly non-linear density distributions close to the critical surface. This means that the ray is so close to the critical surface and so steep that the parabolic switching does not work anymore and the ray crosses the critical surface, which is an unphysical scenario. Then the code goes another step back (from $(\mathbf{r}_{0},\mathbf{v}_{0})$ to $(\mathbf{r}_{-1},\mathbf{v}_{-1})$), which is stored in memory and performs a linear reflection following Snell's law. This condition only occurs for high-frequencies where rays are nearly straight lines and thus the algorithm's precision is not affected. For more details on the ray tracing algorithm and relevant derivations see \cite{Benkevitch:10}.}


\subsection{{Radio Intensity Calculation}}\label{subsec:intensity}

The intensity, $I_\nu$, for each pixel in the image is calculated by integrating the emissivity, $e_\nu$, along the ray, passing through the center of the pixel. The integral can be calculated either along a) the actual refracted (curved - see Sections \ref{subsec:refimpl}, \ref{subsec:raytracing}) path of each ray, or b) the line-of-sight (LOS) ignoring the refraction. The detailed equilibrium principle as applied to the radiation-matter interaction relates the emissivity, $e_\nu$, for each specific radiation mechanism to the product of the irradiation multiplied by the absorption coefficient, $\kappa_\nu$, so that:
\begin{equation}\label{eq:integral}
I_\nu=\int B_\nu(T) \kappa_v \mathrm{d}s,
\end{equation}
with $B_\nu(t)$ being the Planckian spectral intensity of blackbody radiation. 
According to, e.g., \citet[][]{Karzas.Latter:61}, for the Bremsstrahlung emission, we have in the limit $h\nu<<k_BT$
\begin{equation}
B_\nu(T)=\frac{2k_BT_e\nu^2}{c^2} {,}
\end{equation}
\begin{equation}\label{eq:kappa}
\kappa_\nu=\frac{\rho^2_ee^6}{\nu^2(k_BT_e)^{3/2}cm_e^{3/2}}<g_{ff}> {,}
\end{equation}
where $k_B$ is the Boltzmann constant, $e$ is the electron charge, $T_e$ is the electron temperature, $v$ is the frequency of the radio-ray considered, $c$ is the speed of light, $m_e$ is the electron mass and $<g_{ff}>$ is the Gaunt factor, which is taken as $<g_{ff}>=10$ here \citep[see Figure 5 in][]{Karzas.Latter:61}. {The radio intensity per unit frequency is calculated by the ray tracing algorithm based on the MHD quantities computed at each predefined time step. In the current implementation, the intensity does not provide feedback to the MHD simulation, i.e. the emitted energy in radio frequencies does not contribute in the energy equation solved by our simulations.}
%
%

{The intensity of both Bremsstrahlung and simplistic emission is the integral along the ray of the spectral intensity multiplied by the absorption coefficient, as indicated by Equation~(\ref{eq:integral}). The simplistic emission mechanism was implemented in an earlier version of the algorithm based on \cite{Benkevitch:10} and is an approximate approach to combining both Bremsstrahlung and plasma coherent emission. Here we have modified it so that its intensity is defined by Equation~(\ref{eq:integral}), similar to the Bremsstrahlung emission. Its spectral intensity and absorption coefficient are given by}
\begin{eqnarray}
\label{eq:simBv}
B_{\nu,s} =1 , \\
\label{eq:simkv}
\kappa_{\nu,s}= \left(\frac{\rho_e}{\rho_{\rm cr}}\right)^2\left[0.5-\left(\frac{\rho_e}{\rho_{\rm cr}}\right)\right]^2 .
\end{eqnarray}
The plasma frequency is proportional to the square root of {the plasma critical} density, {and is given by}
\begin{equation}
\label{eq:fp}
f_p=\frac{\omega_p}{2\pi}=\frac{1}{2\pi}\sqrt[]{\frac{4\pi e^2 \rho_{cr}}{m_e}} ,
\end{equation}
{in Hz.} {Both the spectral intensity and the absorption coefficient are dimensionless (Equations (\ref{eq:simBv}) and (\ref{eq:simkv})) and thus the intensity of the simplistic radio emission is given in arbitrary units.}
It accounts for the quadratic dependence of Bremsstrahlung emission on the plasma density similar to Equation~(\ref{eq:kappa}). However, unlike the Bremsstrahlung spectrum, it does not account for temperature dependence and employs a modified spectrum over frequencies that peaks at the fundamental ($f_p$) and harmonic ($2f_p$) frequency, thus resembling coherent plasma emission in that aspect. {The simplistic mechanism is an artificial mechanism that was implemented as a zero order approximation to the plasma emission using MHD quantities. In the current implementation, the simplistic mechanism imitates the plasma emission in the sense that they both peak at the plasma frequency and the harmonic similarly \citep[e.g.][]{Zaitsev.Stepanov:83,Bastian.etal:98}, but its intensity and spectrum are consistent with the Bremsstrahlung emission.}

\section{Results} \label{sec:results}

\subsection{{Code Testing and Validation}}\label{subsec:test}

\begin{figure}
\begin{center}
\includegraphics[width=0.47\textwidth]{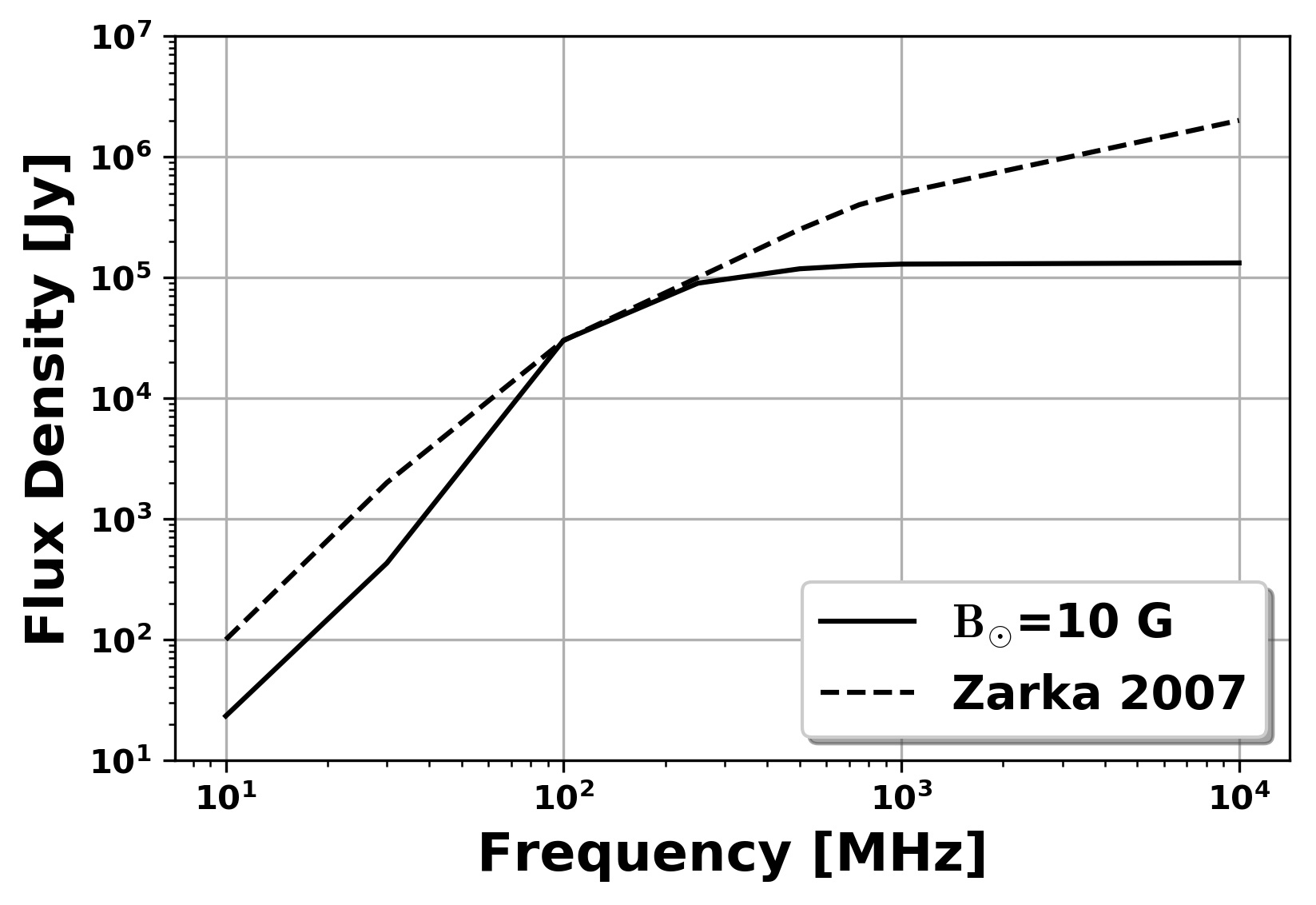}
\caption{{The synthetic radio flux in [$Jy$] (solid line) as a function of frequency as computed by our simulations assuming a simple solar magnetic field of a $10$ G dipole. In the same plot, we illustrate the observed quiet Sun radio flux (dashed line) adapted from \cite{Zarka:07}. Both fluxes are assumed to be observed from the Earth (at a distance of 1~AU).}}
\label{fig:codever}
\end{center}
\end{figure}

{As a first test and in order to verify our simulation, ray tracing, and synthetic imaging algorithm, we ran a steady-state simulation for the purely theoretical case of a solar dipolar magnetic field. The results of that simulation are presented in Figure~\ref{fig:codever}. The strength of the solar magnetic dipole is 10 G and we estimate the flux density of the theoretical case over a wide range of frequencies from 10 MHz to 10 GHz. Our results are compared with the quiet Sun flux density as presented in Fig.~1 of \cite{Zarka:07} and references therein. In both cases, i.e. \cite{Zarka:07} and our simulation, the same trends are being captured. More specifically, the flux density of the radio emission initially increases with increasing frequency up to a few hundreds of MHz. Then the radio flux density starts forming a plateau around GHz frequencies. From this experiment, we conclude that in the absence of active regions, our synthetic imaging tool is able to reproduce flux densities that are in reasonable qualitative agreement with quiet Sun conditions. There is a discrepancy in the high frequency regime due to the fact that in our simulation there are no active regions (pure dipole), which are expected to provide extra radio density flux in higher frequencies \citep{Lee.etal:99}. For that reason, the synthesized density flux initially increases with frequency following the flux density of the Sun as observed from the Earth, reaches a maximum value, and then creates a plateau. For this test we used the Bremsstrahlung emission implementation in our algorithm, which is an indication that the synthetic flux density of the free-free emission is a fairly good approximation of the quiet Sun flux in the range of frequencies examined here.}

\subsection{Observational Test at High Frequency}
\label{s:solar}

\begin{figure*}
\begin{center}
\includegraphics[width=\textwidth]{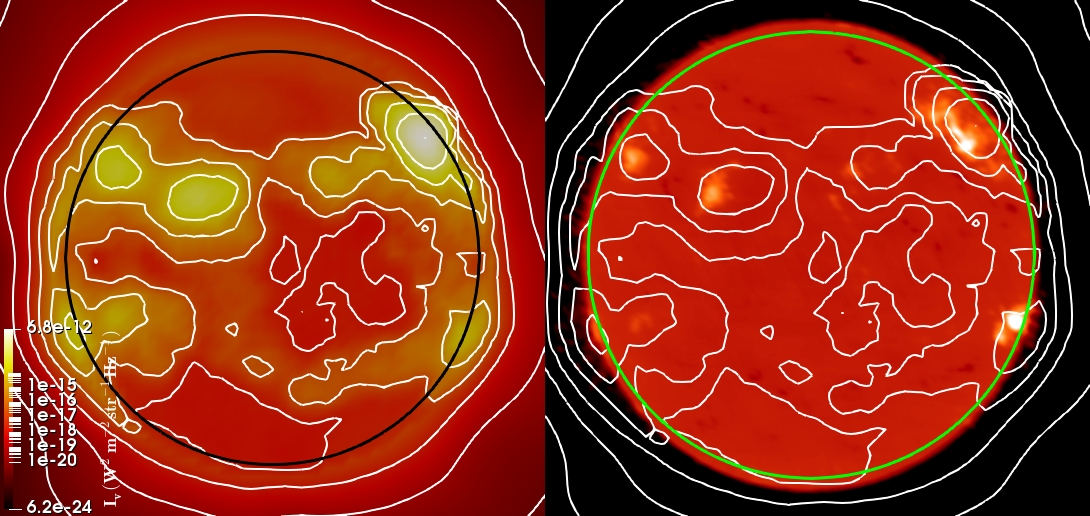}
\caption{Comparison between {synthetic Bremsstrahlung and observed} emission for the full solar disk. Both images correspond to March 7, 2011. Left: Bremsstrahlung radiation as captured in our CR2107 simulation {at 17 GHz}. Right: {Observations from} Nobeyama {Radioheliograph} at 17GHz with the contours corresponding to the simulation overplotted.}
\label{fig:obs}
\end{center}
\end{figure*}

Solar radio observations have been available for a number of years in different frequencies for quiet full sun \citep{Takano.etal:97,Oberoi.etal:11,Mohan.Oberoi:17} and flaring conditions \citep[e.g.][]{Kundu.etal:09,Kochanov.etal:13,Chen.etal:15}. We have tested our model against solar radio observations at MHz and GHz frequencies. In Figure~\ref{fig:obs}, we compare predictions of the {Bremsstrahlung} radio emission from the solar coronal and wind model with observations for Carrington rotation CR2107 March 2011 from the Nobeyama Radioheliograph\footnote{Nobeyama radioheliograph public data can be found at \url{http://solar.nro.nao.ac.jp/norh/}.} at the higher radio frequency of 17 GHz where refraction effects are not expected to be important (see Section~\ref{subsec:refraction} below). {The average radio intensity calculated from our synthetic image at 17 GHz (see Fig.~\ref{fig:obs}) is $6.8\times 10^{-16}$ $\mathrm{Wm^{-2} Hz^{-1} str^{-1}}$ and given that the stereoangle of the Sun as observed from the Earth is $6\times 10^{-5}$ str, the radio flux density becomes then $\sim 10^{-20}$ $\mathrm{Wm^{-2} Hz^{-1}}$. This value captured by our synthetic radio imaging tool matches typical solar observed values by Nobeyama that range from 100 -- 1000 solar flux units (SFU), i.e. $10^{-20}-10^{-19}$ $\mathrm{Wm^{-2} Hz^{-1}}$.}

Our simulated results match the basic features captured by the observations, such as the structures around the solar limb and some features at the center of the disk. Nevertheless, there are fine-scale structures at the solar disk that we do not recover with our simulations. This is expected since the \textit{GONG} magnetogram that we are using as input for our simulations has a relatively low angular resolution of about a degree and is constructed from the monthly averaged (synoptic) map, while the high frequency Nobeyama observations are daily and thus more dynamic and can capture finer radio structures. {The 17 GHz high frequency radio emission mainly originates from the low atmosphere and at these frequencies refraction is expected to play only a very minor role. Thus, the good agreement of our results with observations from Nobeyama serve as a verification that our treatment of the low atmosphere is rather satisfactory and that our ray-tracing algorithm behaves correctly in the high frequency limit case.}


\subsection{{The Low Frequency Regime}}

\begin{figure}
\begin{center}
\includegraphics[width=0.45\textwidth]{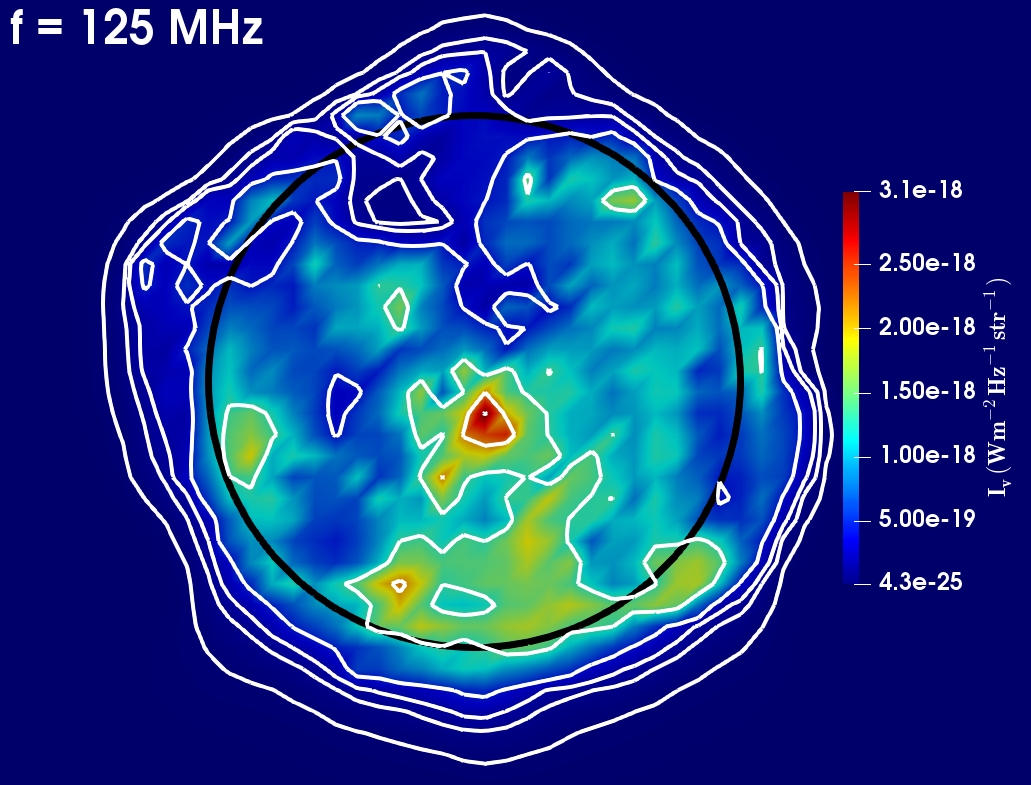}
\caption{{We illustrate the radio intensity map at a frequency of 125 MHz for CR 2156, corresponding to October - November 2014. Our results show a good first order agreement with the observations by MWA in \cite{Mohan.Oberoi:17}.}}
\label{fig:lowf}
\end{center}
\end{figure}

{In Figure~\ref{fig:lowf}, we show the radio synthetic intensity map as calculated by our simulation in the low frequency regime and in particular for a frequency of 125 MHz. Recently, \cite{Mohan.Oberoi:17} used Murchison Widefield Array\footnote{The Murchison Widefield Array (MWA) is a low frequency radio telescope in Western Australia operating at 80 to 300 MHz (\url{http://www.mwatelescope.org}).} (MWA) to obtain low frequency radio observations of the Sun. Refraction is playing a major role in the low frequency regime. As shown in Figure~\ref{fig:lowf}, the solar disk appears more extended in low frequencies due to the critical surface being at a higher altitude. Please note that only our synthetic results are illustrated in Figure~\ref{fig:lowf}. Our results show a good first order agreement with the observations illustrated in Figure 6 of \cite{Mohan.Oberoi:17}. The synthetic image constructed based on our simulation results has captured the most intense feature at the center of the solar disk, as shown in their Figure~6. 
However, there are a few apparent morphological differences between our synthetic image and the MWA observations presented therein. As is illustrated in the different panels of Figure~6 in \cite{Mohan.Oberoi:17} the observed morphological structures in the range of two hundred MHz are very dynamical and change a lot from one second to the next one. We are using synoptic magnetograms, that are time-averaged over the time span of an entire Carrington Rotation.  The simulation in Figure~\ref{fig:lowf} therefore captures a monthly average distribution of the radio features on the solar disk, while the MWA observations of \cite{Mohan.Oberoi:17} also capture their dynamic behavior.}

{Another important source of discrepancy between the MWA observations and our synthetic image at 125 MHz is the effect of scattering, that we do not include in the current version of the ray tracing algorithm. While refraction plays an important role in the low frequency regime, scattering is also expected to be important in that part of the spectrum. However, this is beyond the scope of this work, which is a demonstration and a first verification of the new capabilities for radio synthetic imaging in BATS-R-US. We are planning to include scattering in a future version of the algorithm.}

 
\subsection{{The Role of Refraction}} \label{subsec:refraction}

{For each ray frequency there is a critical surface that forms at some altitude above the photosphere where the local MHD plasma density becomes equal to the critical density $\rho=\rho_{cr}$. That critical surface is the last layer (different for each frequency $\omega$) that is accessible by that particular ray. Electromagnetic rays of that frequency $\omega$ cannot penetrate in the region below that critical surface and as they approach they undergo strong refraction and finally bend away, as shown in Figures 7--9 in \cite{Benkevitch:10}. The critical surface is closer to the photosphere for higher frequency rays \citep[see Fig. 9,][]{Benkevitch:10}. The region between the critical surface and the photosphere is not accessible to rays with frequencies $\omega <\omega_p$. The refractive index $n$ decreases gradually as a ray approaches the critical surface, until below it becomes imaginary once it crosses that surface. Exactly at the critical surface the refractive index becomes zero $n=0$, the ray frequency equals the local plasma frequency $\omega=\omega_p$, and the plasma density is equal to the critical density for that ray $\rho=\rho_{cr}$. Our ray tracing algorithm captures this behavior by calculating the refractive index and estimating the critical surface for each ray. It then computes the curved trajectory for each ray, as explained in section \ref{sec:method} and \cite{Benkevitch:10}.}

\begin{figure*}
\begin{center}
\includegraphics[width=\textwidth]{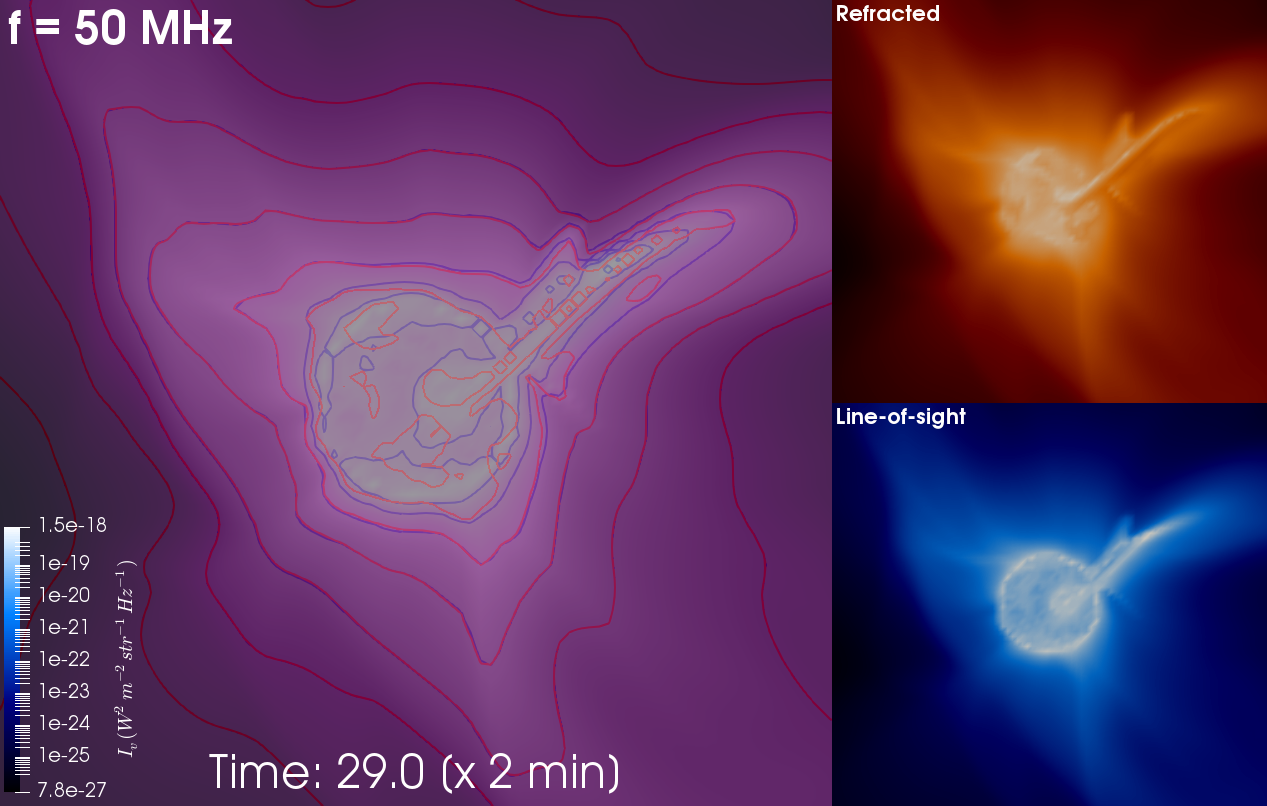}
\caption{Radio synthetic view at 50MHz along the actual curved (top right) and line-of-sight straight (lower right) paths {$\sim$ 1 hour} after the CME ejection time for CR2107. We also show both scenarios by overlaying them (left panel) together with their respective contours to help quantify their differences. There are clear morphological differences between the two scenarios in both the solar disk and the CME regions (see text for details).}
\label{fig:refnoref}
\end{center}
\end{figure*}

In Figure~\ref{fig:refnoref}, we illustrate the importance of the refraction by quantifying the radio intensity using the refracted curved ray paths (top right panel) \textit{versus} calculating the radio intensity along straight lines (bottom right panel). In the left large panel we have overlaid the two cases by decreasing the opacity of the refracted image. The difference in the size and intensity of the solar disk and the CME become evident in Figure~\ref{fig:refnoref} by visual inspection. More specifically, from this preliminary study, it becomes evident that the main morphological differences between the line-of-sight and refracted trajectory integration are focused on two regions, namely at the solar disk and at the center of the CME. {In particular, even though both synthetic images come from the same MHD solution, the solar disk has a high intensity ring around its edge for the line-of-sight integration and the two images have different distributions of individual radio sources at the solar disk body and the CME core.} The contours, illustrated for both cases in Figure~\ref{fig:refnoref}, highlight the differences in the radio intensity of the two scenarios that vary significantly at the bright central part of the CME, with quite different, dynamically changing, features being picked up in each case. {Finally, we provide a movie of the full simulation time series similar to \ref{fig:refnoref} as online material.}


\subsection{CME dynamics}
\label{s:dynamics}

\begin{figure*}
\begin{center}
\includegraphics[width=\textwidth]{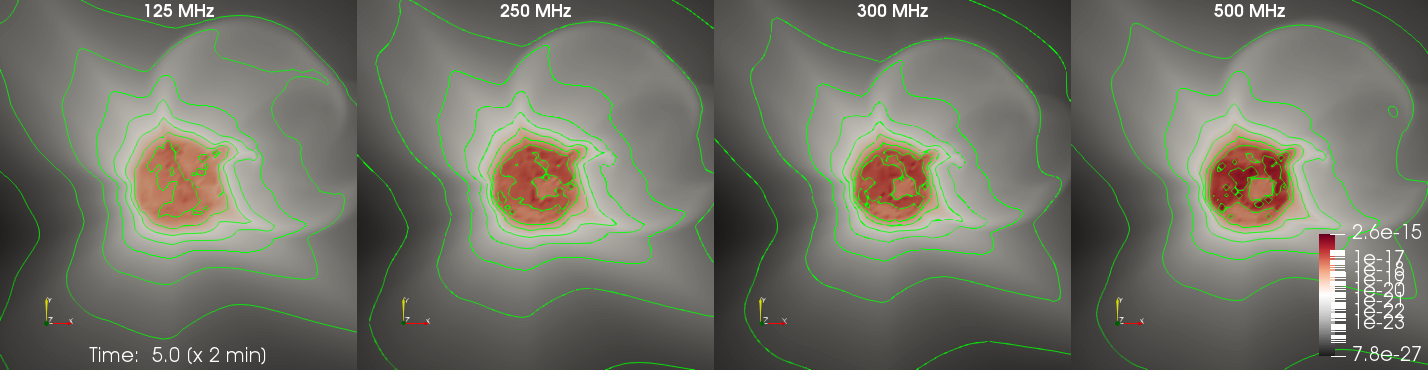}
\includegraphics[width=\textwidth]{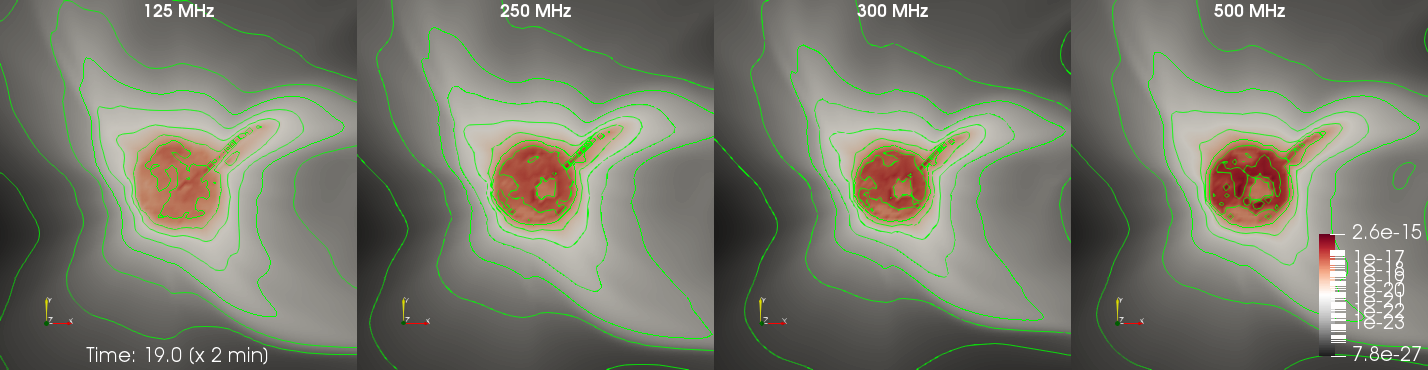}
\caption{Images of our simulated Bremsstrahlung radio-loud CME at frequencies of 125 MHz, 250 MHz, 300 MHz and 500 MHz superimposed with ten logarithmically spaced intensity contours 10 (top) and 38 (bottom) minutes  after the CME initiation time. {This type of CME modeling can be compared with the radio CME observations presented in \citet{Bastian.etal:01}.}}
\label{fig:cmeobs}
\end{center}
\end{figure*}

Currently, radio observational imaging capabilities do not allow for dynamic CME studies as detailed and streamlined as quiet solar disk ones.
Even though imaging eruptive filaments is relatively easy due to their high density \citep[e.g.][]{Gopalswamy.etal:03,Kundu.etal:04}, capturing a CME in Bremsstrahlung radio emission is difficult as they become very diffuse quickly after eruption. Nevertheless, CMEs have been imaged in radio during their eruption phase \citep{Gopalswamy.Kundu:92,Bastian.etal:01,Ramesh.etal:03,Demoulin.etal:2012,Zucca.etal:18}.


{Unfortunately, there is no Nobeyama high-frequency flare observation available 
corresponding to the simulated CME during CR2107. However, we can still compare qualitatively our simulation results and synthesized images with radio observations of past CMEs.} In figure \ref{fig:cmeobs}, we show four different synthetic images at frequencies of 125 MHz, 250 MHz, 300 MHz and 500 MHz, after 10 (best CME volume illustration) and 38 minutes (best illustration of point-source peaks) from the CME initiation time. {These Bremsstrahlung CME images can be compared with the radio CME observed by \citet[][see Figure 3]{Bastian.etal:01}, even though those authors attribute that radio emission to non-thermal synchrotron radiation.} More specifically, the top part of the CME is less radio intense, while the footpoints appear to be more radio-loud. Furthermore, the higher the frequency, the smaller the size of the radio-loud lower parts of the CME legs appear. {Unfortunately, there are no synoptic magnetograms from \textit{GONG} before September 2006 for us to drive our simulation and compare with the event captured in \citet{Bastian.etal:01}.
{The dominant emission captured in their observations is not the free-free emission, which is implemented in our code, but rather the synchrotron radiation \citep{Bastian.etal:01}. However, the synchrotron intensity strongly depends on the magnetic field, and our 3D MHD simulation can provide the thermal density and the vector magnetic field self-consistently.}

Figure \ref{fig:cmeobs} and videos illustrating the CME development and provided as on-line material demonstrate that our radio synthetic images of the Bremsstrahlung emission capture an intense large-scale radio wave propagating behind the CME in the Sun-ward direction and expanding towards the solar disk limbs as the CME expands. Interestingly, as pointed out by \citet{Casini.etal:17}, Bremsstrahlung and EUV emission have the same square density dependence. Similarly to the large-scale radio wave captured here, large-scale EUV waves associated with CMEs and flares are systematically observed close to the solar surface \citep[e.g.][]{Ireland.etal:18}.

\begin{figure*}
\begin{center}
\includegraphics[width=\textwidth]{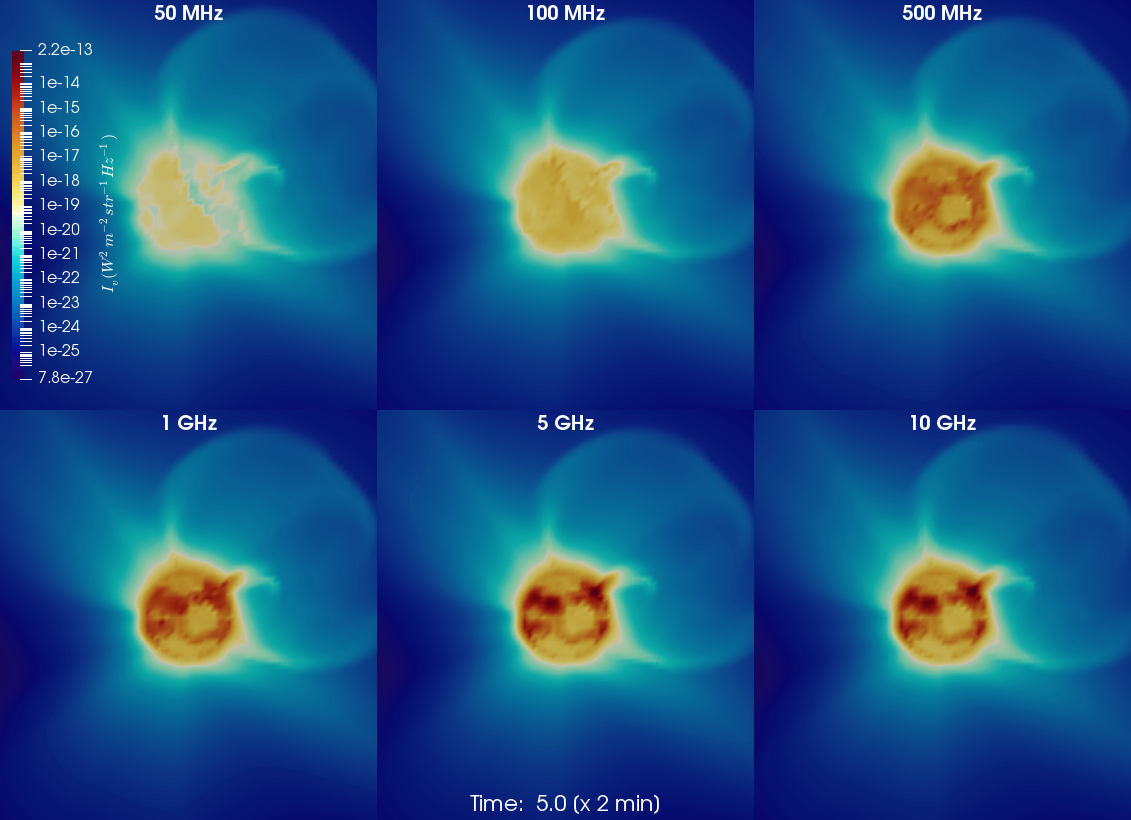}
\caption{Bremsstrahlung multi-frequency radio synthetic views at 50 MHz, 100 MHz, 500 MHz, 1 GHz, 5 GHz and 10 GHz. Higher frequencies probe deeper into the solar atmosphere and capture the finer scales of the magnetic field such as ARs. The radio intensity of the solar disk and the CME is higher for higher frequencies (also see text).}
\label{fig:Lee}
\end{center}
\end{figure*}

Radio-rays with higher frequencies penetrate deeper into the solar atmosphere and thus have access to stronger magnetic fields and can resolve finer magnetic field structures on the solar surface, as explained in \cite{Lee.etal:99}. In figure \ref{fig:Lee}, we show a multi-panel radio image of the solar surface with the CME captured in our simulations in six different frequencies: 50 MHz, 100 MHz, 500 MHz, 1 GHz, 5 GHz and 10 GHz. The CME appears overall dimmer for low frequencies with the CME cavity being more easily distinguished at the lowest frequency. As we increase the ray frequency, the radio signal from the central part of the CME appears stronger in comparison to the entire CME volume. There is a dim spherical shell at the CME location captured in all frequencies. Furthermore, the radio emission of the entire extended solar "surface" has a lower intensity for lower frequencies.

These results are consistent with \cite{Lee.etal:99}, i.e. the solar "surface" appears larger for lower frequencies as they are refracted at  higher altitude, while higher frequencies capture the fine scales on the solar "surface" with active regions (ARs) emitting in higher intensity. {Essentially, this is a combination of the variable intensity of the refraction effects with frequency and the fact that the MHD quantities are dependent on each other in self-consistent MHD solutions. {Even for the Bremsstrahlung} emission there is an indirect dependence of the radio intensity on the local magnetic fields through the MHD density and temperature variability. Based on this fact \cite{Casini.etal:17} presented a new method using Bremsstrahlung measurements to quantify solar magnetic fields.} {We would like to note here that currently there is no radioheliograph that can provide imaging over such a large field of view and wide frequency range as shown in Figure \ref{fig:Lee}. Thus, results similar to those demonstrated in Figure \ref{fig:Lee}, will be useful for guiding the next generation solar radio telescopes, such as the Frequency Agile Solar Radio telescope (\textit{FASR\footnote{\url{http://www.fasr.org}}}) and the Chinese Spectral Radioheliograph (\textit{CSRH\footnote{\url{http://english.nao.cas.cn/Research2015/Facilities2015/Telescopes2015/201701/t20170120_173590.html}}}) anticipated in the future.}

\subsection{{Bremsstrahlung \textit{versus} Simplistic implementation}}
\label{subsec:nonthermal}

Bremsstrahlung emission is consistently implemented in our code for the first time. In this section, we compare the newly implemented Bremsstrahlung emission with the previous prescription, which ignores the temperature dependence and has a spectrum with peaks at the fundamental and second harmonic plasma frequencies. The simplistic version of Figure~\ref{fig:Lee} is illustrated in Figure~\ref{fig:simple}. Note here that the units in the two figures are different, with the Bremsstrahlung intensity given in SI irradiance units, while the simplistic intensity is in dimensionless arbitrary units. The logarithmic scale can act as an easy visual way to compare the structures, signal intensity and range of variations between the two mechanisms.

\begin{figure*}
\begin{center}
\includegraphics[width=\textwidth]{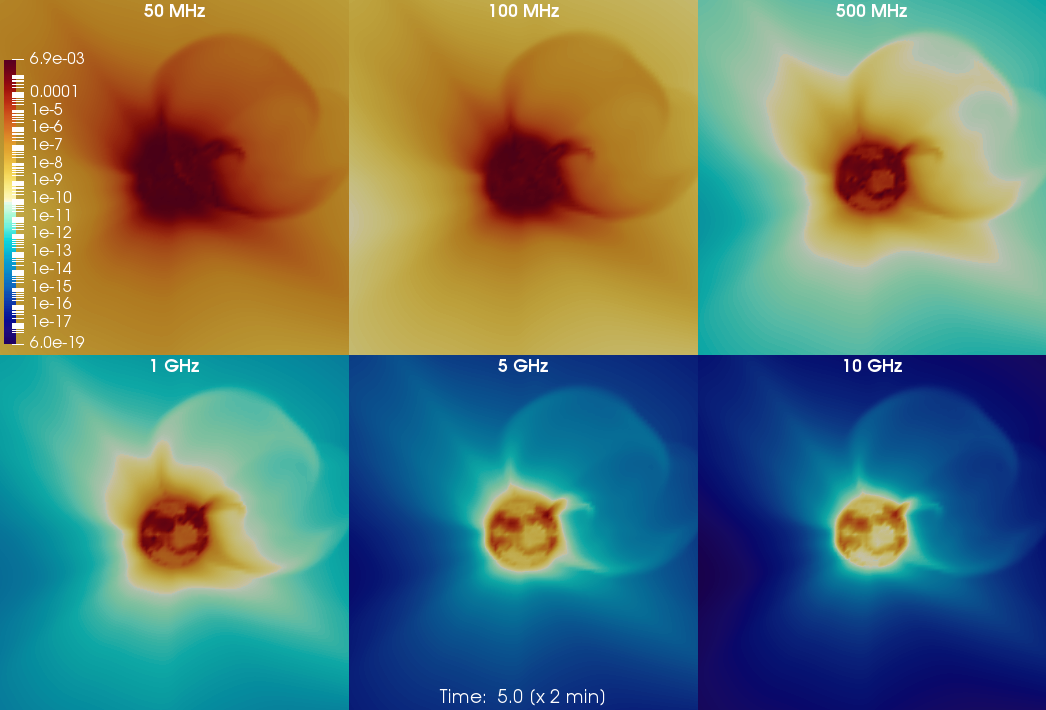}
\caption{Similar to Figure \ref{fig:Lee}, but here illustrating the radio images  capturing the simplistic radio emission at the same frequencies: 50MHZ, 100MHz, 500MHz, 1GHz, 5GHz and 10GHz. Note that even though higher frequencies have again access to larger magnetic fields and progressively resolve them better, the picture is reverse compared to Figure \ref{fig:Lee}, since the overall emission intensity gradually decreases with frequency.}
\label{fig:simple}
\end{center}
\end{figure*}

Morphologically, both mechanisms capture the same structures throughout the range of frequencies considered. This is due the fact that the computed radio intensity in both cases has the same strong dependence on density. Nevertheless, there is a number of differences between the simplistic and the Bremsstrahlung emission. Most importantly, the entire radio-loud extended solar disk, the background, and the CME appear louder in lower frequencies for the simplistic case. As we move toward higher frequencies, the background, the CME, and the parts of the solar disk without ARs are becoming dimmer by several orders of magnitude. In the simplistic case, the emission of the CME is changing more drastically between different frequencies, contrary to the Bremsstrahlung case, where the CME radio intensity is almost constant for the range of frequencies illustrated. This is due to the fact that while flat spectra are expected for Bremsstrahlung emission (see Section ~\ref{sec:discussion}), the simplistic mechanism, following the plasma emission, peaks at the plasma frequency and second harmonic \citep[e.g.][]{Bastian.etal:98}. The plasma frequency is given by $\omega_p\propto 9000\sqrt[]{n_e}$, so for typical coronal densities, e.g. of the order of $n_e\approx10^8\mathrm{cm^{-3}}$, the simplistic mechanism is expected to peak at frequencies of approximately $\approx$100-200MHz.

\section{Discussion}
\label{sec:discussion}

\subsection{{The Importance of Bremsstrahlung Emission}}

{The intensity of the Bremsstrahlung emission  only depends on macroscopic plasma quantities that are computed by our MHD simulations and as such it is the least computational demanding mechanism and the natural first step for a numerical implementation. Even though Bremsstrahlung emission is not considered to be the dominant mechanism during solar energetic phenomena, its implementation is not just a numerical test. }

As discussed by \cite{Casini.etal:17}, observations of the Bremsstrahlung emission probe solar surface magnetic field distributions and evolution. They can thus be complimentary to existing observational methods that measure the photospheric magnetic field of the Sun, as it is currently difficult to directly measure the magnetic field in the outer solar atmosphere. Moreover, the Bremsstrahlung emission has the same square density dependence ($\rho^2$) as the extreme ultraviolet (EUV) brightness. EUV observations have been a valuable tool for solar dynamics. Similarly to EUV observations, the Bremsstrahlung emission captures density variations and has the potential of providing important insights and being complimentary to current and future EUV observational and numerical capabilities. For example, from our simulations a density wave was captured just behind the CME propagating towards the solar surface and expanding outwards. Similar structures linked to CMEs are extensively studied using EUV solar observations and are called Moreton waves \citep[e.g.][]{Harra.etal:11,Veronig.etal:11,Francile.etal:16}.

\subsection{{Beyond the Bremsstrahlung Emission}}\label{subsec:mechanisms}

\begin{figure*}
\begin{center}
\includegraphics[width=0.4\textwidth]{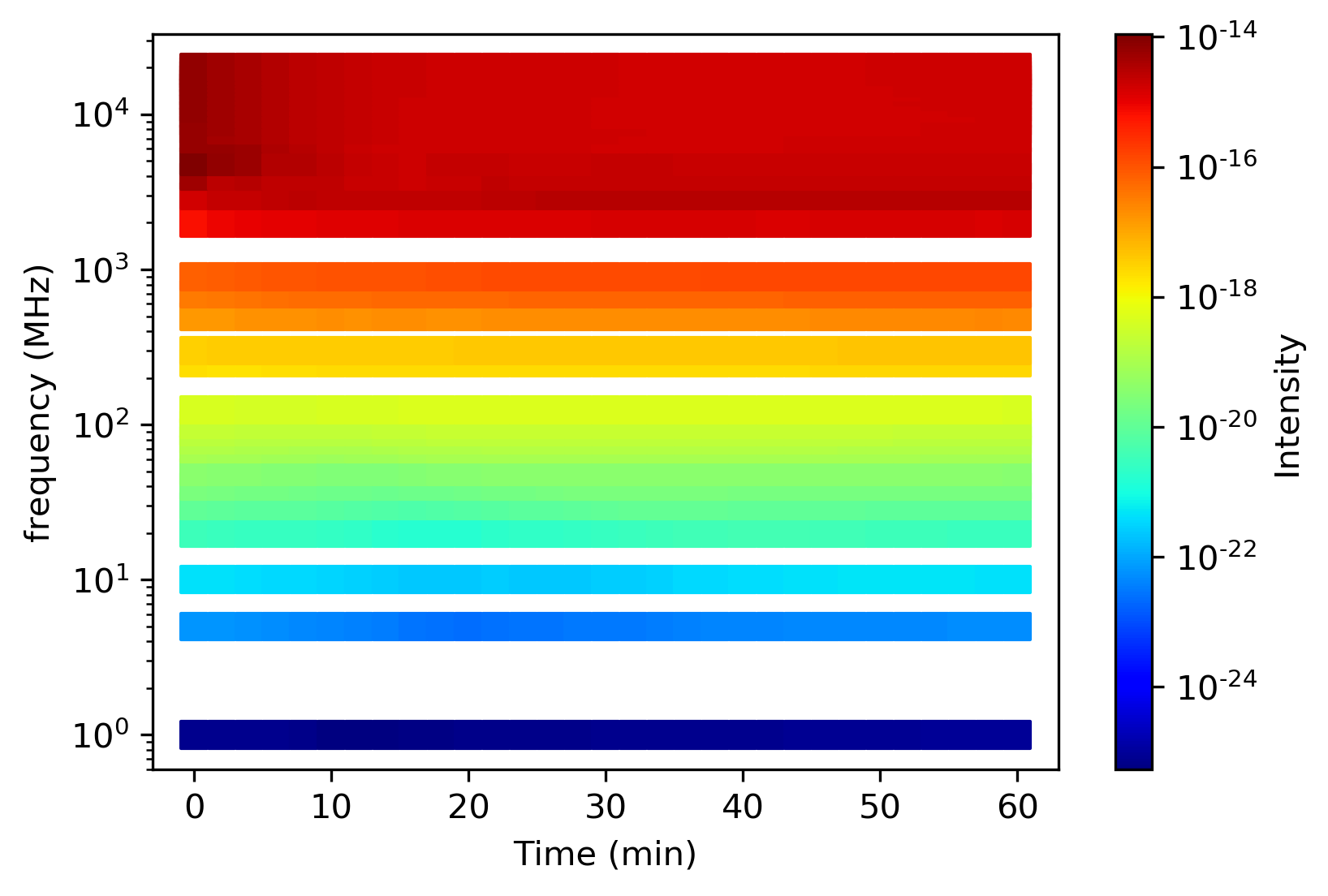}
\includegraphics[width=0.52\textwidth]{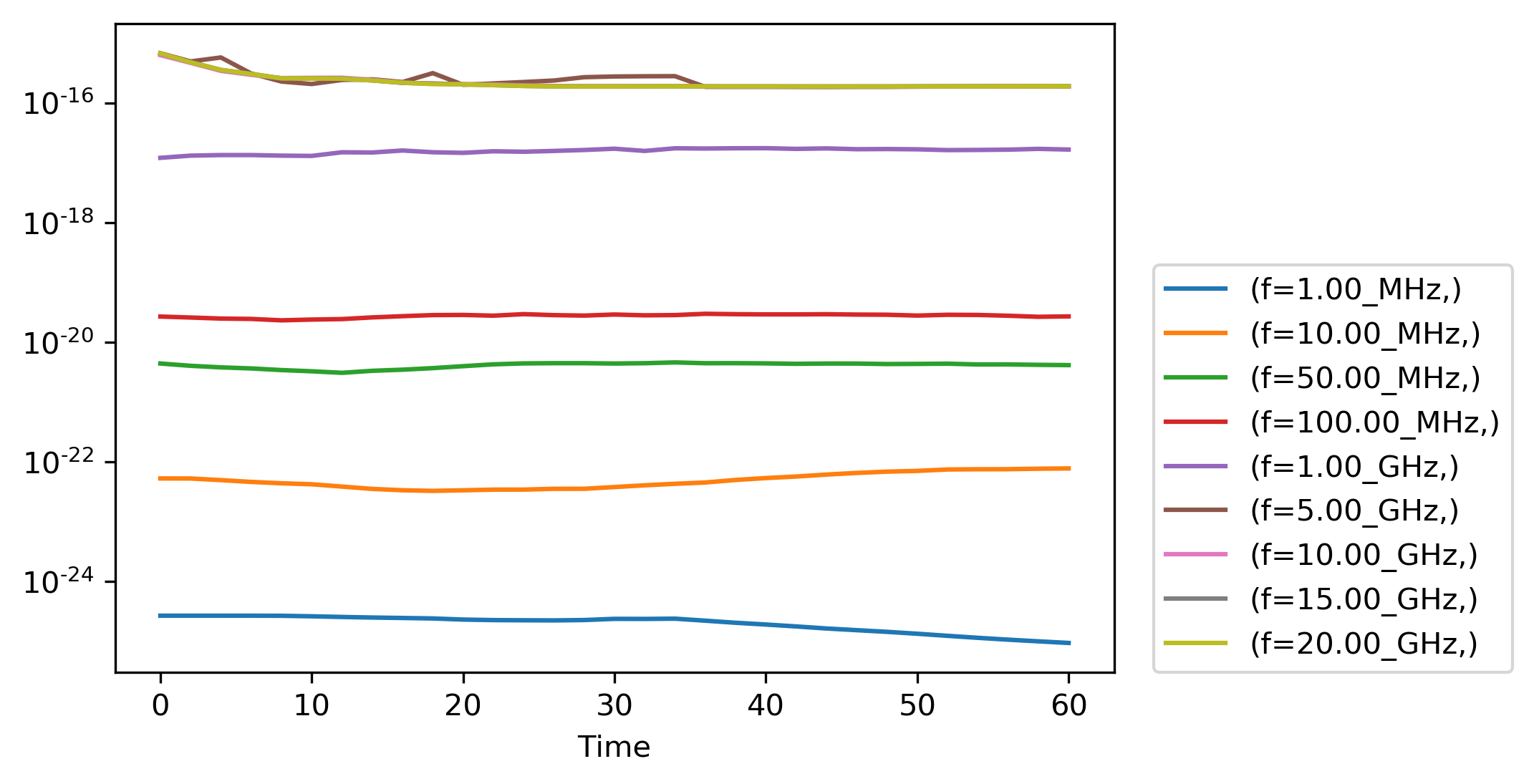}
\caption{{Left panel: Dynamic spectrum for radio synthetic images of the Bremsstrahlung emission. Right panel: We plot a few characteristic frequencies of the integrated synthetic images over time to show how the average intensity varies with frequency. The Bremsstrahlung intensity illustrated here is given in SI.}}
\label{fig:spectra}
\end{center}
\end{figure*}

We illustrate in Figure~\ref{fig:spectra} the dynamic radio spectrum for our synthetic CME event computed by integrating the intensity of successive synthetic radio images made for the same viewing angle as in Figures \ref{fig:refnoref}--\ref{fig:Lee}. Image snapshots were made every two minutes from CME initiation up to 1 hour after the flux-rope eruption. {Each radio synthetic image corresponds to a specific predefined frequency. In principle, any frequency from a few KHz up to the order of THz values can be used in our algorithm. A finite number of frequencies can be used per simulation for computational and also practical reasons, as we cannot create an infinite number of output files. More specifically, 36 frequencies from 1 MHz to 20 GHz were chosen and used for the purposes of this study, as shown in the dynamic spectrum of Figure~\ref{fig:spectra}. The gaps in the left panel of Figure~\ref{fig:spectra} correspond to frequencies for which we did not produce a synthetic image. In the right panel of Figure~\ref{fig:spectra} we illustrate 9 selected frequencies spread throughout the 5 orders of magnitude covered in our frequency range. Illustrating only a limited number of integrated fluxes as curves allows for an easier interpretation of our results.}

The dynamic spectrum visualization serves as an extra test to verify that the Bremsstrahlung spectrum is captured properly at radio frequencies. {Note that the integrated fluxes corresponding to frequencies larger than 5 GHz are indistinguishable, as shown in the right panel of Figure~\ref{fig:spectra}. This is due to the fact that the main part of the high frequency emission comes from the active regions in combination with the limited resolution of the \textit(GONG) magnetogram used in our simulation, as discussed in Section \ref{s:solar}. In other words, we cannnot resolve the active region topology any further by increasing the frequency above some threshold. Minor temporal variations in the integrated fluxes are captured in the right panel of Figure~\ref{fig:spectra}. This is mainly due to refraction combined with the CME evolution described in our simulation. The dynamic evolution of the transient makes the corona a highly non-uniform and time-dependent medium. The calculation of the actual curved ray paths per frequency depends on the MHD parameters of the plasma region it travels through, as discussed in Section \ref{subsec:refraction}. Higher frequencies will penetrate deeper in the solar atmosphere and thus different parts of the computational volume will be used for the integration of different fluxes at different times during the evolution of the simulation. We are mainly recovering the flat spectrum of the Bremsstrahlung emission mechanism. However, rays with different frequencies have a different response to the same plasma parameters, which explains the weak intensity variations captured.} 

It is worth noting here that this capability will become more physically interesting once burst-like, i.e. plasma coherent emission mechanisms, are included in our radio emission computation. {More specifically,  once the generating emission mechanisms are included in a future version of this tool, synthetic radio bursts are expected to show characteristic frequency drifts in their dynamic spectra that reflect the changing plasma parameters of the emitting source during the burst.}

CMEs are observed directly at radio frequencies in our solar system due emission mechanisms associated with non-thermal phenomena. As discussed in Sections {\ref{subsec:intensity}, the simplistic mechanism currently available does not suffice for studies of non-thermal effects, such as radio bursts. Coherent plasma emission is essential for capturing the shock waves at the CME shock front tied to Type II radio bursts. More rigorous treatments of these non-thermal mechanisms need to be included in our tool and we will do so in a future study. 

{The next least computationally demanding radio emission mechanism in terms of implementation for an MHD code is gyrosynchrotron emission \citep{Dulk:85}. The intensity of the gyrosynchrotron emission depends on the macroscopic MHD plasma parameters, similar to Bremsstrahlung emission. More specifically, gyrosynchrotron emission scales as $\rho T^\alpha B^\beta$, with $T$ the temperature, $B$ the magnetic field strength and $\alpha, \beta > 1$. However, even for this mechanism we need to start deviating from the single fluid MHD solution and assume a pitch angle distribution for the electrons. The implementation of plasma emission requires more robust particle prescriptions in our simulations. The coupling of MHD and kinetic physics will increase considerably the computational cost of our numerical experiments.}

\subsection{{Implications for Stellar Research}}

Observations in radio frequencies can also play an important role in the fast developing field of exoplanetery discoveries and can help in providing a deeper understanding of the environments surrounding planetary systems. Radio emission can be generated by the interaction of the stellar wind plasma with magnetospheres of magnetized exoplanets \citep[e.g.][]{Zarka:07,Lazio:07,Grismeier:07,Vidotto:15,Burkhart.Loeb:17,Llama.etal:18,Turnpenney.etal:18}. Unfortunately, this emission from star-planet interaction is undetectable from the Earth due to the terrestrial ionosphere having a lower cut-off frequency of $\sim 10$~MHz \citep{Davies:69,Yeh.Liu:82}, while the exoplanetary radio aurora frequencies lie below this limit, as discussed in \cite{Burkhart.Loeb:17}. 

For habitability considerations, it is important to understand the activity characteristics of a host star. Stellar winds and CMEs remain extremely difficult to detect on Sun-like stars and incontrovertible stellar CME observations are sorely needed \citep[e.g.][]{Moschou.etal:17}. Type II radio bursts produced by shocks generated in the interaction of CMEs and the ambient coronal environment \citep[e.g.][]{Gopalswamy.etal:01,Gopalswamy.etal:09,Liu.etal:13,Cunha-Silva.etal:15} are currently one of the most promising observational methods for detecting shock waves from stellar CMEs. While stellar CMEs have not yet been definitively observed \citep[see][and references therein]{Leitzinger.etal:11,Crosley.etal:16,Crosley.Osten:18a,Crosley.Osten:18b}, and the spatial resolution of stellar radio observations cannot yet approach the resolution attainable for the Sun, detection of analogous events in other stars can provide key observational constraints for CME behavior and parameters \citep{Crosley.etal:16}.

The work presented here provides a means of probing the emission levels and radio properties of CMEs in stellar environments, whose properties can differ from the solar case by orders of magnitude in terms of magnetic activity. Such predictions are prerequisite for planning resource-demanding observational campaigns and for the design and planning of future observational facilities.

\section{Conclusions} \label{sec:conclusions}

We have presented a radio synthetic imaging tool that is integrated in a state-of-the-art three-dimensional MHD simulation of the solar corona and inner heliosphere. {The Bremsstrahlung emission is calculated along the actual curved ray paths in the three dimensional space.} In this study we focus on Bremsstrahlung emission, which has the same dependency on density ($\propto \rho^2$) as EUV and X-ray line-dominated emission and is thus able to capture density variability. The radio counterpart of EUV waves forming behind a CME and propagating towards the Sun were captured. {Results obtained with the radio synthetic imaging algorithm presented in this paper can provide guidance for the next generation radio missions.} It is the first time that refraction has been so systematically examined in a realistic simulation of the solar corona. 

We are planning to extend our calculations to non-thermal radio emission mechanisms in the future and include scattering in our algorithm. Meanwhile, case-studies of Bremsstrahlung emission will help better understand the insights into coronal physics in different plasma regimes that this mechanism can provide.

Even though here we mainly focus on the Sun, we would like to emphasize the scientific importance of radio synthetic imaging capabilities in combination with state-of-the-art MHD for astrophysics in general. This development should help in disentangling different complex radio emission processes, as it allows for complete freedom in terms of including or excluding different mechanisms and allowing for focused studies of each mechanism in physically interesting setups.



\acknowledgments
We thank an anonymous referee for her/his useful comments and suggestions. SPM and OC were supported by NASA Living with a Star grant number NNX16AC11G. JJD was supported by NASA contract NAS8-03060 to the {\it Chandra X-ray Center}. JDAG was supported by Chandra grants AR4-15000X and GO5-16021X. CG was supported by Chandra grants GO7-18017X and GO5-16021X. Simulation results were obtained using the Space Weather Modeling Framework, developed by the Center for Space Environment Modeling, at the University of Michigan with funding support from NASA ESS, NASA ESTO-CT, NSF KDI, and DoD MURI. Simulations were performed on NASA's PLEIADES super-computer under award HEC-SMD-17-1190.

\bibliographystyle{yahapj}
\bibliography{references}


\end{document}